\documentstyle[aps,preprint]{revtex}

\tighten
\def \cT {{\cal T}}

\def \cG {{\cal G}}

\def \cT {{\cal T}}

\begin{document}
\title{comment on: " Even-odd behavior of conductance in monatomic sodium wires" }
\date{\today}
\author{}
\maketitle

\narrowtext

Based on numerical {\bf LDA} calculation in their recent letter
Sim et al. report an even-odd behavior of the conductance as a
function of the number of atoms in monatomic sodium
wires.\cite{Sim} For a string of up to $N=5$ sodium atoms with
ends connected to leads with mirror symmetry, they found the
conductance to have the value $\cG_{0}=2e^{2}/h$ for odd $N$,
while for even $N$ it is smaller and sensitive to the wire
structure. We here give a simple analytic proof of this result,
and demonstrate that such an even-odd behavior exists in more
general cases. We also argue that this parity effect holds for
larger $N$ as long as $N$ is finite.

 We model the wire as a string
of  $N$ ${\bf s}$-orbital ($Na$, $Au$, $Ag$, $Cu$)
atoms\cite{Yeyati} with its ends connected to leads, by the
following tight-binding hamiltonian
\begin{equation}
H=
\sum\limits_{i=1}^{N}\epsilon_{s}a_{is}^{+}a_{is}+\sum%
\limits_{i=1}^{N-1}(t_{i}a_{is}^{+}a_{i+1s}+H.c.)+\sum\limits_{k,r=L,R}%
\epsilon
_{kr}b_{kr}^{+}b_{kr}+\sum%
\limits_{k}(V^{L}_{ks}b_{kL}^{+}a_{1s}+V^{R}_{ks}b_{kR}^{+}a_{Ns}+H.c.),
\end{equation}
where $a_{is}$$(b_{kr})$ is the annihilation operator of an ${\bf
s}$-orbital electron at atomic site $i$ (lead $r$), and the other
terms have their usual meaning.  The Landauer-B\"{u}ttiker
conductance is determined by the transmission probability at the
Fermi level, $\cG=\cG_{0}\cT(E_F)$. Considering the atomic sites
to the right of site $1$ as part of the right lead, the Keldysh
formalism yields for the transmission probability the expression,

\begin{eqnarray}
&&  \nonumber \\
\cT(E_{F}) &=&\frac{-2\Gamma
^{L}|t_{1}|^{2}ImG_{2R}^{r}}{(E_{F}-\epsilon_s
-|t_{1}|^{2}ReG_{2R}^{r})^{2}+(\Gamma
_{L}-2|t_{1}|^{2}ImG_{2R}^{r})^{2}/4}.
\end{eqnarray}
where
\begin{eqnarray*}
\Gamma ^{L/R} &=&\sum\limits_{k}2\pi |V^{L/R}_{ks}|^{2}\delta
(\epsilon
-\epsilon _{kL/R}), \\
G_{iR}^{r}(E_{F}) &=&[g_{is}^{r}(E_{F}))^{-1}-\mid t_{i}\mid
^{2}G_{i+1,R}^{r}(E_{F})]^{-1},\hspace{1cm}i=2,3,\cdots N-1   \\
G_{NR}^{r}(E_{F}) &=&[g_{NN}^{r}(E_{F}))^{-1}+\frac{i}{2}\Gamma
^{R}]^{-1}
 \\
g_{is}^{r}(E_{F}) &=&(E_{F}-\epsilon_{s}+i0^+
)^{-1}.\hspace{1cm}i=1,2,\cdots N
\end{eqnarray*}
At resonance ($E_{F}=\epsilon_s )$ the real part of all retarded
Green functions becomes zero and the first term in the denominator
of Eq. (2) vanishes. Complete transmission through the dot lattice
is then obtained if
\begin{equation}
\Gamma ^{L}=-2|t_{1}|^{2}ImG_{2R}^{r},
\end{equation}
where now

\begin{eqnarray}
ImG_{2R}^{r} &=&-\frac{\Gamma ^{R}}{2}|\frac{t_{3}t_{5}\cdots t_{N-2}}{%
t_{2}t_{4}\cdots t_{N-1}}|^{2};\hspace{0.2cm}N\hspace{0.1cm}odd \\
ImG_{2R}^{r} &=&-\frac{2}{\Gamma ^{R}}|\frac{t_{3}t_{5}\cdots t_{N-1}}{%
t_{2}t_{4}\cdots t_{N-2}}|^{2};\hspace{0.2cm}N\hspace{0.1cm}even.
\end{eqnarray}
The condition thus becomes

\begin{eqnarray}
|\frac{t_{1}t_{3}\cdots t_{N-2}}{t_{2}t_{4}\cdots t_{N-1}}|^{2} &=&\frac{%
\Gamma ^{L}}{\Gamma ^{R}};\hspace{0.8cm}N\hspace{0.1cm}odd \\
|\frac{t_{1}t_{3}\cdots t_{N-1}}{t_{2}t_{4}\cdots t_{N-2}}|^{2} &=&\frac{%
\Gamma ^{L}\Gamma ^{R}}{4};\hspace{0.3cm}N\hspace{0.1cm}even.
\end{eqnarray}
Equation (6) states that for a sample with an odd number of atoms
and mirror symmetry ($\Gamma^{L}=\Gamma
^{R},t_{1}=t_{N-1},t_{2}=t_{N-2},etc.)$ complete transmission is
automatically satisfied, yielding $\cG=2e^{2}/h.$ This is not the
case when $N$ is even, however, as is apparent from the structure
of Eq. (7). For a wire with deformations perpendicular to its
length the even-odd character of conductance is preserved since
the normal deformation just changes the inter-site couplings
symmetrically. One also finds that the conductance quantization
for even $N$ is more sensitive to the wire structure.

The above results are valid for any finite N. Also, they appear
not to arise from the electron-electron interaction since our
derivation ignores such coupling. If the latter is added, however,
we expect them to hold as well since the on-site Coulomb
interaction  introduces a self-energy term $\Sigma _{e-e}$ in the
Green's function of each site.  The influence of electron-electron
interactions is then just to shift and split the resonance
position \cite{Langreth} and Eq. (2) can also be formally used in
their presence, with the formal replacement
$g_{is}^{r}=(E_{F}-\epsilon_s -Re\Sigma _{e-e})^{-1}$ and the new
resonance condition $\epsilon_s =E_{F}-Re\Sigma _{e-e}$.

\vspace{30pt} \noindent This work was supported in part by a
Catedra Presidencial en Ciencias and FONDECYT 1990425 (Chile), and
NSF grant No. 53112-0810 of Hunan Normal University (China).

\vspace{30pt}

\noindent Z. Y. Zeng$^{1,2}$ and F. Claro$^{1}$

$\phantom{x}$\vspace{-11pt}

$^{1}${\small Facultad de F\'isica, Pontificia Universidad Cat\'olica de
Chile, Casilla 306, Santiago 22, Chile } \newline
$^{2}${\small Department of Physics, Hunan Normal University, Changsha
410081, China}\newline


\end{document}